\documentclass[]{article}
\usepackage{graphicx}
\usepackage[latin1]{inputenc}
\usepackage{amsmath}
\usepackage{comment}

\usepackage{xcolor}
\usepackage[urlcolor=blue]{hyperref}      
\hypersetup{
    colorlinks = true,                    
    citecolor = {blue},
    linkcolor = {purple},
           }

\title{\bf Emotions as abstract evaluation criteria
   in biological and artificial intelligences} 

\author{Claudius Gros \\
Institute for Theoretical Physics,  \\
Goethe University Frankfurt am Main, Germany}

\begin{document}
\maketitle

\begin{abstract}
Biological as well as advanced artificial intelligences 
(AIs) need to decide which goals to pursue. We review 
nature's solution to the time allocation problem, 
which is based on a continuously readjusted categorical 
weighting mechanism we experience introspectively as 
emotions. One observes phylogenetically that the 
available number of emotional states increases hand 
in hand with the cognitive capabilities of animals and 
that raising levels of intelligence entail
ever larger sets of behavioral options. Our ability 
to experience a multitude of potentially conflicting 
feelings is in this view not a leftover of a more 
primitive heritage, but a generic mechanism for attributing 
values to behavioral options that can not be specified at 
birth. In this view, emotions are essential for
understanding the mind.

For concreteness, we propose and discuss a framework
which mimics emotions on a functional level. Based
on time allocation via emotional stationarity (TAES),
emotions are implemented as abstract criteria, such 
as satisfaction, challenge and boredom, which serve 
to evaluate activities that have been carried out. 
The resulting timeline of experienced emotions is 
compared with the `character' of the agent, which 
is defined in terms of a preferred distribution of 
emotional states. The long-term goal of the agent, 
to align experience with character, is achieved 
by optimizing the frequency for selecting individual 
tasks. Upon optimization, the statistics of emotion 
experience becomes stationary.

\end{abstract}

\section{Introduction}

Humans draw their motivations from short- and long
term objectives evolving continuously with new
experiences \cite{huang2014selfish}. Here we argue 
that this strategy is dictated in particular by the 
fact that the amount of information an agent disposes 
about the present and the future state of the world is 
severely constraint, given that forecasting is 
intrinsically limited by successively stronger 
complexity barriers \cite{gros2012pushing}. 
Corresponding limitations hold for the time available 
for decision making and for the computational 
power of the supporting hard- or wetware 
\cite{zenon2019information,lieder2020resource}, 
independently of whether the acting agent 
is synthetic or biological. A corollary of this
observation is that the time allocation problem,
which goals to pursue consecutively, cannot be 
solved by brute force computation. Instead, nature 
disposed us with an emotional control system. 
It is argued, in consequence, that an improved 
understanding of the functional role of
emotions is essential for theories of the mind.

Emotions have emerged in the last decades as 
indispensable preconditions for higher cognition 
\cite{panksepp2004affective,gros2010cognition}.
It has been pointed out, in particular, that the 
core task of an emotional response is not direct causation 
of the type ``fleeing because one is afraid'', but the
induction of cognitive feedback, anticipation and reflection
\cite{baumeister2007emotion}. Being afraid will in general 
not result in a direct behavioral reaction, but in the
allocation of cognitive resources to the danger at hand.
If there is chance, it is better to attempt to solve an 
existing problem. It has been shown in this context that 
emotional and cognitive processes form a tight feedback 
system in terms of emotional priming of cognition 
\cite{beeler2014kinder}, and cognitive control of 
emotions \cite{ochsner2005cognitive}. Cognitive emotion 
regulation \cite{inzlicht2015emotional}, such as the 
attempt to restrain one's desire for unhealthy food, 
is present to such a stage \cite{cutuli2014cognitive}, 
that it can be regarded as a defining characteristics 
of our species. The advanced cognitive capabilities that are 
paramount to efficiently pursue a given goal, like 
winning a game of Go, will hence leave their imprints 
also on the cogno-emotional feedback loop 
\cite{miller2018happily}. 

On a neuronal level it has been observed 
\cite{pessoa2008relationship} that the 
classical characterization of brain regions
as `affective' and `cognitive' is misleading 
\cite{pessoa2019embracing}. The reason is that 
complex cogno-emotional behaviors are based rather 
on dynamic coalitions of networks of brain areas 
\cite{pessoa2018understanding}, than on the 
specific activation of a single structure,
such as the amygdala \cite{phelps2006emotion}. The same 
holds for the neural representations of the cognitive control 
mechanisms regulating emotional responses, which are found 
to be distributed within a network of lateral frontal, temporal,
and parietal regions \cite{morawetz2016neural}.

The interconnection of cognitive and emotional brain 
states suggests a corresponding dual basis for decision making
\cite{lerner2015emotion}. Logical reasoning would then
be responsible to analyze alternative choices, with the 
outcome of the different choices being encoded affectively
as values \cite{reimann2010somatic}. An equivalent
statement holds for the weighting of the associated risks 
\cite{panno2013emotion}, in particular when it comes to 
long-term, viz strategic decision taking \cite{gilkey2010emotional}. 
One feels good if a specific outlook is positive and certain, 
and uneasy otherwise. The consequence is hence that
intelligence is needed to rationalize decision options,
but that intelligence alone, if existing in terms of a 
pure logical apparatus, cannot solve the time allocation 
problem. Logic is not enough to decide which long-term
goals to pursue one after another. It has been argued, in 
analogy, that self-control is intrinsically not a purified 
rational, but a value-based decision process \cite{berkman2017self}.

The picture emerging is that the brain uses deductive 
and other types of reasoning for the analysis of behavioral 
options, see \cite{shepard1971mental,johnson2010mental,papo2015can},
and emotional states for the weighting of the consequences.
One observes that distinct types and combination of 
emotional states, like anger, envy, trust, satisfaction,
etc, are attributed to specific types of behavioral options
\cite{pfister2008multiplicity,schlosser2013feeling}, which
implies that the number of emotional states an agent 
needs for the weighting of its options increases hand 
in hand with the complexity of decision making. A 
consequence of this interrelation between emotional
and cognitive complexity shows up in daily life,
to give an example, when it comes to handle adversity 
effectively, which has 
been shown to benefit from emotional 
diversity \cite{grossmann2019wise}.

As an intriguing corollary of here discussed scenario 
we note that an hypothetical artificial intelligence
of human and transhuman level should dispose,
as a conditio sine qua non, of a palette of emotional 
states containing ever finer shades of states. Synthetic 
emotions would be in this framework equivalent to human 
emotions on a functional level, but not necessarily in
terms of corresponding qualia. The reason 
for the increased emotional sensibility of 
advanced AIs is that the number of available 
weighting categories has to match the increased number 
of behavioral and cognitive options accompanying high 
intellectual capabilities. The challenge to guarantee 
the long term dynamical stability of the corresponding 
feedback loop between motivations, goal selection and 
introspective cognitive analysis, viz the task to
control advanced AIs, will hence increase 
in complexity with raising intelligence levels.

It has been observed that one obtains an
improved level of understanding if working 
principles for the brain are not only formulated, 
but also implemented algorithmically 
\cite{cauwenberghs2013reverse,hawkins2017special}. 
In this spirit we will present, after discussing
the relation between emotions and feelings, a simple
but operational cogno-emotional framework. The goal is
to show that the concept of emotional stationarity
allows to select varying timelines of subsequent 
tasks. Cognitive abilities are in this view 
important to complete a given task, with emotional
values being responsible for task selection in
first place. 

For our discussion, a multi-gaming environment will 
be used as a reference application. Within this 
protocol, agents have two qualitatively different 
tasks. First to select what to do, the time 
allocation problem at its core, and then to 
complete the selected job. Having finished a game, 
say Go or chess, agents will evaluate the acquired 
experience emotionally, with the timeline of 
experiences shaping in turn the decision process 
of what to do next.

\subsection{Life-long utility maximization}

From the perspective of Darwinian evolution,
life-long utility maximization is directly
proportional to the number of offsprings.
For most humans, this is nowadays not the goal,
if it ever has been. As mentioned before, we
presume here that the aim is instead to align 
character and experiences and that this
process is mediated by emotions. Support
from neuropsychological research will be
discussed further below. As a matter of
principle one could imagine, alternatively,
that life-long utility maximization is
achieved by calculating at any moment the
optimal course of action, while discounting 
the entirety of future rewards. For a
variety of reasons this is however not
possible, even if exceedingly large
computational powers would be at one's
disposal.

In machine learning, the scaling of performance
with the amount of dedicated resources has
been investigated. Within the domain of language
processing, it has been shown that the performance
of deep-learning architectures scales as a power-law 
of any of the three primary scarce resources, time, 
model- and training-data size 
\cite{kaplan2020scaling}, if not bottlenecked by
one of the others. It is good news that 
exponentially larger amounts of resources are
not needed to boost the performance within a 
well specified application domain. The computational
demands needed by top deep architectures increases 
in contrast exponentially with time
\cite{agneeswaran2020computational},
at a rate that outpasses Moore's law by far
\cite{greifman2020efficient}. From the perspective
of complex systems theory \cite{gros2015complex},
this is not a surprise, given that state-of-the-art 
machine learning architectures are applied to 
increasingly complex problems and domains. For 
example, when forecasting horizons are extended, the 
intrinsically chaotic nature of most complex
systems demands exponentially increasing computing
times. This difficulty has been termed
`complexity barrier' \cite{gros2012pushing}.

In societies, complexity barriers arise in addition
from the need to predict and to interpret the behavior 
of the other members, which is of course a reciprocal 
task. Indeed, the `social brain hypothesis' 
\cite{dunbar2009social} states that a core evolutionary 
driver for the development of our brains has been 
the need to deal with the complexity of human social 
systems, with the latter evolving in parallel with
increasing brain sizes. From an evolutionary perspective, 
a cognitive intra-species arms race with progressively 
increasing computational resources leads to a 
'red-queen phenomenon' \cite{dieckmann1995evolutionary}, 
namely that it takes `all the running to stay in place'.
These two factors, the eventual occurrence of an 
intra-societal cognitive arms race, and the
intrinsic complexity of the environment, makes it
impossible to predict the future via brute force
computations, in particular for the purpose to 
maximize life-long utilities. It is to be seen
if an analogous line of arguments holds for
societies of advanced artificial intelligences.


\section{Experiencing emotions as feelings}

Before delving further into the analysis of the 
functional role of emotions, we take a step back 
and ask a deceivingly simple question. Why do we
have feelings in first place?

At any given point of time, a myriad of neural,
chemical and electrical processes take place in 
our brains. For the overwhelming part, consciously 
we are however not aware of what our supporting wetware
is doing \cite{van2012unconscious,dehaene2014toward}. 
In contrast, we are able to experience as 
feelings \cite{wang2016neuromodulation} the 
class of processes corresponding to emotional
states \cite{colibazzi2010neural}. Why then has evolution
developed neural circuits allowing our brain to
experience feelings? The alternative would be that
the functional role of emotions would be performed
by neural processes we could not register consciously.
In this case we would be akin to what has been called 
at times a `zombie' \cite{koch2001zombie}, viz a
human-like being which is not aware of its emotional 
drives \cite{winkielman2004unconscious}. A zombie
would just go for the food, when hungry, without being
able to restrain itself. Defined as such, zombies 
are not able to close the cogno-emotional
loop \cite{inzlicht2015emotional,miller2018happily}, 
lacking the capability to control emotions cognitively.
The human condition is based, in contrast, on 
emotional control as a defining trait. This is the 
underlying reason why people with reduced impulse 
control skills, e.g.\ when intoxicated or drunk,
are considered more often than not to be less 
accountable for their doings 
\cite{penney2012impulse}, at times to the extent 
that they are exempted from criminal 
liability.

It is presently not fully settled how we are able to 
experience the feelings arising in conjunction to 
emotional brain states. A series of experiments 
point in this regard to a feedback loop involving 
the response of the body \cite{levenson2014autonomic}.
The prospect would be that emotional brain states invoke
bodily reactions, like an increased heart-beating rate, 
that would be transmitted back as `gut feelings' to the 
brain via propriosensation \cite{nummenmaa2014bodily}, 
that is through visceral and other peripheral 
sensors \cite{kreibig2010autonomic}. Of interest is 
here that the cortical region responsible for 
channeling the afferent propriosensation, the anterior
insular cortex, is fully developed only in higher apes and 
hence phylogenetically young \cite{craig2009you}. Animals 
unable of self recognition seem to lack the spindle-shaped 
economo neurons characteristic of the anterior insular cortex.
Deactivating the brain regions allowing us to sense
our own body would bring us hence one step closer 
to losing the ability to experience emotional states as
feelings \cite{de2010mind}. Given that evolution has
taken care to equip us with feelings, they must 
improve Darwinian fitness, entailing hence important 
functionalities.

A vast number of studies has shown that emotional 
processes regulate the attributing of values 
to stimuli \cite{cardinal2002emotion} and that 
they bind conceptual information through affective
meaning \cite{roy2012ventromedial}. Being able to 
experience these brain processes consciously in 
terms of feelings is therefore a necessary condition 
for the conscious control of the brain's value system. 
Feelings are in this view the keystone closing the 
feedback circle between cognitive information 
processing and the emotional value system. Our 
preferences and disinclinations would be fully 
subconscious, and not controllable, if we would 
not be able to perceive them introspectively as 
feelings. This line of arguments, which relates 
the introspective experience of emotional states 
to the ability to be aware of one's own value system, 
is in our view likely to be the rationale for the 
phylogenetic emergence of feelings.

\subsection{Emotions in non-human animals}

Emotions are not unique to humans 
\cite{ledoux2012evolution}, but functional states 
of the nervous system that can be studied across 
species \cite{anderson2014framework,perry2017studying}.
An emerging consensus in the field is that animal and 
human emotions have functional equivalent roles with 
regard to decision making \cite{mendl2020animal}.
Going down the phylogenetic tree, the decreasing 
complexity of the nervous system entails however 
that the range of possible affective states narrows 
progressively. For example, it has been observed that 
fish may appraise environmental stimuli cognitively 
\cite{cerqueira2017cognitive}, that flies can 
express anxiety \cite{mohammad2016ancient}, and 
that the decision-making behavior in bumblebees 
seems analogous to optimism in humans 
\cite{perry2016unexpected}, at least on an 
operational level \cite{baracchi2017insects}. It 
is however difficult to imagine that a fly could 
take pride in her doings, or experience any other of 
the myriads of human emotional state that obtain
their significance from social context.

Humans are set apart from the other animals populating 
earth not only because of their cognitive capabilities, 
but also because of their ability to experience not 
just a few, but a vast variety of emotional states. 
Studies of heartbeat perception tasks have found, as 
discussed in the previous section, that the substrate 
for subjective feeling states is provided by a 
phylogenetically young brain region, the anterior 
insular cortex \cite{craig2009you}, via a representation 
of evoked visceral responses \cite{critchley2004neural}. 
The anterior insular cortex plays however not only a prominent 
role in the experience of emotions, but also in the value 
attribution system, enabling behavioral flexibility 
\cite{kolling2016value,ebitz2016dorsal}. During 
decision-making, the dorsal anterior cingulate cortex 
is thought to regulate the tradeoff between exploring 
alternative choices, and persistence. A related 
viewpoint links the dorsal anterior cingulate cortex to
the allocation of computational resources to decision 
making \cite{shenhav2016dorsal}.
From a somewhat philosophical point of view one may hence
ask whether it is a coincidence, a caprice of nature 
\cite{gros2009emotions}, to say, that humans are at the 
same time the most intellectual and the most emotional 
species \cite{mendl2011animal,maximino2015non}. It may 
alternatively be a conditio sine qua non. Higher cognitive 
powers would be in the latter case dependent on an 
equally evolved emotional system
\cite{vitay2011neuroscientific}.

\section{An exemplary cogno-emotional framework}

In the following we provide an example for a 
bare-bone cogno-emotional architecture. The 
aim is to demonstrate that our proposed
concept, emotions as abstract evaluation 
criteria, is valid in the sense that it
can be implemented algorithmically. No 
claims are made that the framework examined, 
TAES (``time allocation via emotional
stationarity''), has direct correspondences
to specific brain states or processes.
For illustrational purposes, an application
scenario from machine learning is used
\cite{rumbell2012emotions,jordan2015machine}. 
Specifically, we discuss a multi-gaming 
environment, viz the case that the agent,
f.i.\ a machine-learing AI, decides on its own 
which game to play next.

\subsection{Multi-gaming environments}

Modern machine-learning algorithms based on
deep-neural nets are able to play
a large variety of distinct games
\cite{schrittwieser2020mastering}, such as Go, 
chess and Starcraft, or console games like Atari. 
We consider a setup where the opponents may be 
either human players that are drawn from a 
standard internet-based matchmaking system, 
standalone competing algorithms, or agents 
participating in a multi-agent challenge 
setup \cite{samvelyan2019starcraft}. Of minor 
relevance to the question at hand is the 
expertise level of the architecture and 
whether game-specific algorithms are used. A 
single generic algorithm \cite{silver2017masteringGo},
such as standard Monte Carlo tree search supplemented 
by a value and policy generating deep network
\cite{silver2017mastering}, 
would do the job. For our purpose, the key 
issue is not the algorithm actually playing,
but the question whether the process determining 
which task to select, viz which game to play
at any given time, is universal. In particular, 
we demand that the task selection process can be 
adapted in a straightforward manner when the
palette of options is enlarged, f.i.\ when the
possibility to connect to a chat room is added.

We stress that the framework introduced here, TAES,
is rudimentary on several levels. A fully developed
cogno-emotional feedback loop is not present, which
is in part because present-day agents are neither 
able to reflect on theirselves, no to reason 
rationally on a basic level. TAES serves however 
as an implementable illustration of emotional 
task selection and evaluation.

\begin{figure}[!t]
\centering
\includegraphics[width=0.75\textwidth]{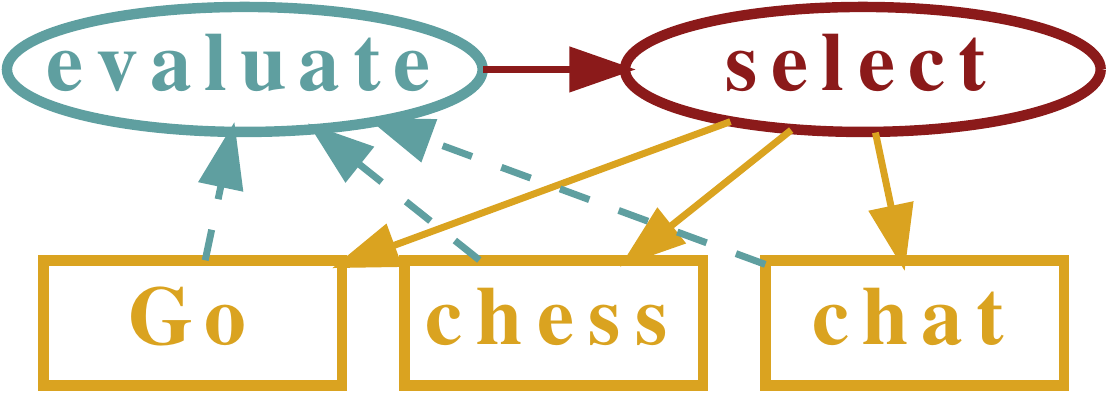}
\caption{Illustration of a general time-allocation framework.
The different options, here to play Go, to play chess and to chat,
are evaluated emotionally once completed, with the evaluation 
results feeding back into the decision what to do next.
TAES, time allocation via emotional stationarity, is a
specification of the general framework.
}
\label{fig_frameworkGeneral}
\end{figure}

\subsection{Emotional evaluation criteria}

In a first step one needs to define the qualia 
of the emotional states and how they are
evaluated, viz the relation of distinct
emotions to experiences. The following
definitions serve as examples.
\begin{itemize}
\item[--] {\sl Satisfaction.} Winning a game raises 
	  the satisfaction level. This could hold 
          in particular for complex games, that is for 
          games that are characterized, f.i., by an
          elevated diversity of game situations.
\item[--] {\sl Challenge.} Certain game statistics may
          characterize a game as challenging. An example
          would be games for which the probability to win
          dropped temporarily precariously low.
\item[--] {\sl Boredom.} Games for which the probability
          to win remains constantly high could be classified 
          as boring or, alternatively, as relaxing. The 
          same holds for overly long games.
\end{itemize}
The key point of above examples is that they can
be implemented algorithmically. Once a task is
performed, which means that the game is played till 
the end, the history of moves can be analyzed 
and the game classified algorithmically along 
above emotional criteria. See 
Fig.~\ref{fig_frameworkGeneral} for an illustration.

The implementation of the evaluation procedure 
depends on the computational framework used. 
Consider f.i.\ the generic deep architectures 
AlphaGo \cite{silver2017mastering} and
AlphaZero \cite{silver2018general},
which consist of layered networks with
two heads, one for the policy and one for
the value, together with a Monte-Carlo tree 
search for valuable game positions. For a
given game position, the value head outputs 
an estimate for the probability to win. 
A game could be classified hence as boring 
when the chance to win, as predicted by the 
value head, remains constantly high, say 
above 70\%. The policy head suggests likewise 
possible high yielding moves, which helps to 
guide the generation of the Monte Carlo search 
tree. A tree characterized by a single 
main stem proposes only a limited number of 
possible good moves. A complex and widely branched 
tree would in contrast be equivalent to a
challenging situation, with larger numbers
of possible moves. An elevated frequency of
complex search trees would classify a game
therefore as challenging. These two examples 
of evaluation criteria abstract from the semantic
content of what the agents is actually doing, a
defining property. They are hence suitable 
to evaluate if any full-information two-player 
game without random components, the application 
domain of AlphaGo and AlphaZero, is boring or 
challenging.

Emotions correspond to value-encoding variables, 
denoted for above example with $S$, $C$ and $B$, 
respectively for satisfaction, challenge and boredom. 
Alternative emotional qualia would be defined 
equivalently. It is important note to keep in mind
that the aim of our framework is to model the 
core functionality of human emotions, but not 
necessarily their affective meanings, which implies 
that it is not mandatory for the evaluation criteria 
to resemble human emotions in terms of their 
respective qualia. The latter is however likely 
to make it easier, f.i., to develop an intuitive 
understanding of emotionally-driven
robotic behavior.  

\begin{figure*}[!t]
\centering
\includegraphics[width=0.75\textwidth]{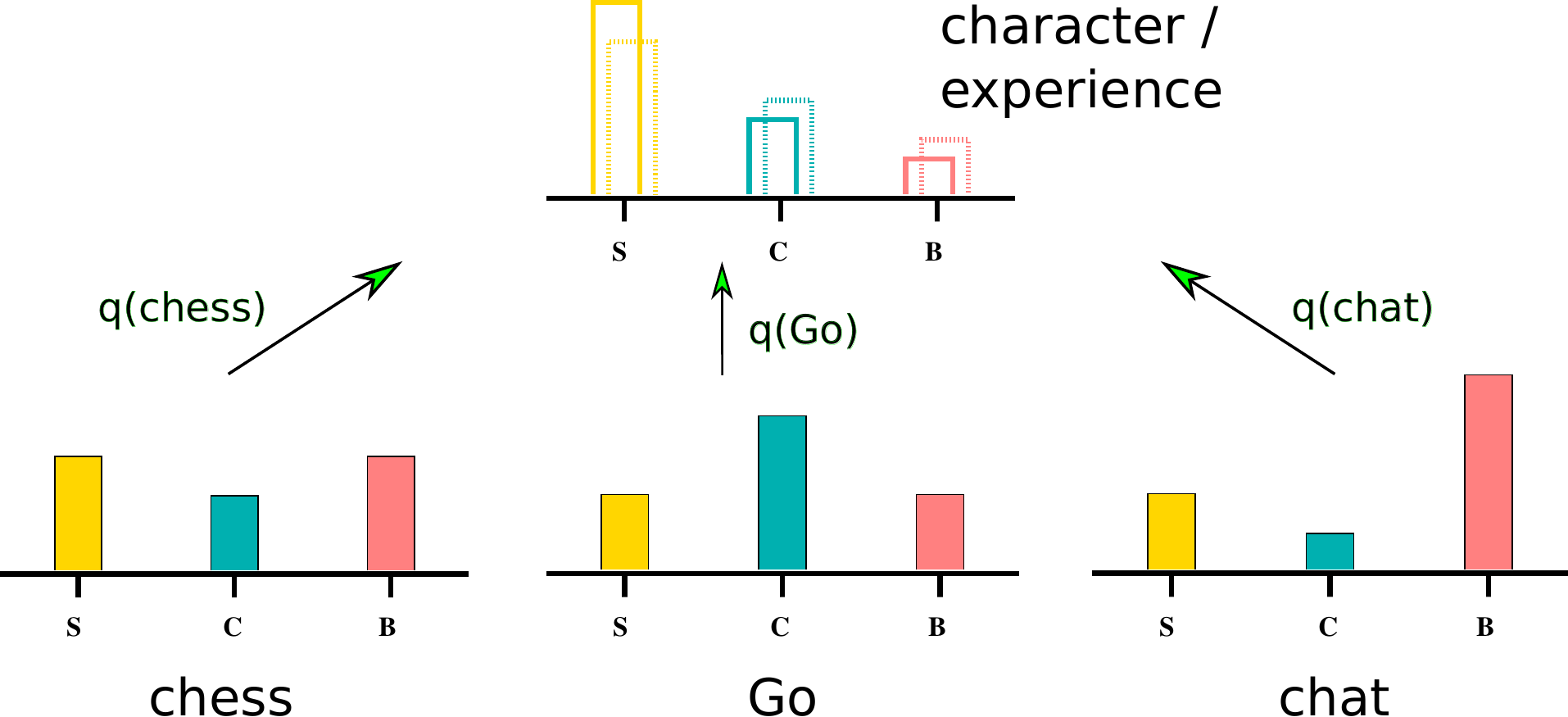}
\caption{Aligning experience with character. Behavioral
options (playing chess, playing Go, joining a chat) are 
evaluated along emotional criteria, such as being satisfying (S), 
challenging (C) or boring (B). The corresponding probability
distributions are superimposed with weights 
$q_\alpha=q(\alpha)$, where 
$\alpha\in\{\mathrm{chess},\mathrm{Go},\mathrm{chat}\}$.
See Eq.~(\ref{E_total}). The goal is to align a predefined
target distribution of emotional states, the character,
with the actual emotional experience. This can be achieved
by optimizing the probabilities $q_\alpha$ to engage in
activity $\alpha$.
}
\label{fig_characterExperience}
\end{figure*}

\subsection{Direct emotional drivings vs.\ emotional priming}

Standard approaches to modeling synthetic 
approaches often assume that emotional state
variables are explicit drivers of actions 
\cite{rodriguez2015computational}, either 
directly or via a set of internal motivations
\cite{velsquez1997modeling}. This means that
a state variable corresponding f.i.\ to being 
'angry' would be activated by specific events,
with the type of triggering stimuli being hard 
coded, viz specified explicitly by the programmer.
Here we are interested in contrast in frameworks 
that are generic, in the sense that behavior is 
only indirectly influenced by emotional states 
\cite{beeler2014kinder}. This implies, as
illustrated in the previous section, that 
emotional evaluation abstracts in its basic
functionality from semantic content.

Within TAES, the agent updates in a first step 
its experience. For every type of activity, 
say when playing Go, the probability that a game 
of this type is challenging, boring or satisfying 
is continuously updated. It could turn out, e.g., that 
Go games are typically more challenging and less boring 
than chess games. Based on this set of data, the 
experience, the next game will be selected with 
the aim to align experience as close as possible 
with the `character' of the agent, which is defined 
in the following.

\subsection{Aligning experience with character}

We define the character $\mathbf{C}_A$ of the 
agent as a preset probability distribution of 
emotional states,
\begin{equation}
\mathbf{C}_A=\big\{P_S,P_C,P_B\big\},
\qquad\quad
P_S+P_C+P_B=1\,,
\label{C_A}
\end{equation}
where $P_S,P_C,P_B\ge 0$ are the target frequencies
to experience a given emotional state. The character
is hence defined as the set of individual
preferences. Agents with large $P_C$\,/\,$P_B$ would 
prefer for example challenging\,/\,boring
situations. The overall objective function of the 
agent is to align experience with its character. 
This means that agent aims to experience satisfying,
challenging and boring situations on the average
with probabilities $P_S$, $P_C$ and $P_B$.

On a basic level, experience can be expressed as
a set of $N$ probability distribution functions,
\begin{equation}
\mathbf{E}^\alpha=\big\{p_S^\alpha,p_C^\alpha,p_B^\alpha\big\},
\qquad\quad
\alpha=1,\dots,N\,,
\label{E_alpha}
\end{equation}
where $N$ is the number of possible activities
(playing Go, chess, connecting to a chat room, ...).
For every option $\alpha$ the agent records,
as illustrated in Fig.~\ref{fig_characterExperience},  
the probability $p^\alpha_i$ for the activity 
to be satisfying/challenging/boring ($i=S/C/B$).
Defining with $q_\alpha$ the likelihood 
to engage in activity $\alpha$, the overall 
experience $\mathbf{E}_A$ is given as
\begin{equation}
\mathbf{E}_A = \sum_\alpha q_\alpha \mathbf{E}^\alpha,
\qquad\quad
\sum_\alpha q_\alpha =1\,,
\label{E_total}
\end{equation}
where the $\mathbf{E}^\alpha$ are defined 
in (\ref{E_alpha}). The global objective, 
to align character $\mathbf{C}_A$ and 
experience $\mathbf{E}_A$, can be achieved 
by minimizing the Kullback-Leibler divergence 
between $\mathbf{C}_A$ and $\mathbf{E}_A$ 
with respect to the $q_\alpha$. This strategy, 
which corresponds to a greedy approach, could 
be supplemented by an explorative component 
allowing to sample new opportunities 
\cite{auer2002using}. Modulo exploration,
an activity $\alpha$ is hence selected with 
probability $q_\alpha$.

TAES is based on aligning two probability 
distribution functions, $\mathbf{E}_A$ and $\mathbf{C}_A$, an 
information-theoretical postulate that has been denoted 
the `stationarity principle' \cite{echeveste2015fisher}
in the context of neuronal learning \cite{trapp2018ei}
and critical brain activity \cite{gros2021devil}.
It states that not the activity as such should
be optimized, but the distribution of activities.
The resulting state is consequently varying in time, but
stationary with respect to its statistical properties.
The underlying principle of the here presented framework
corresponds to `time allocation via emotional 
stationarity' (TAES). Within this approach, the 
character of the agent serves as a guiding functional,
a stochastic implementation of the principle of guided 
self-organization \cite{gros2014generating}.

\subsection{Motivational drives}

Up to now we considered purely stochastic 
decision making, namely that activities are 
selected probabilistically, as determined 
by the selection probabilities $q_\alpha$. An 
interesting extension are deterministic components 
corresponding to emotional drives. Considering 
finite time spans, we denote with $p_i(N_a)$ the 
relative number of times that emotion $i=S,\,C,\,B,\,...$ 
has been experienced over the course of the 
last $N_a$ activities. Ideally, the trailing 
averages $p_i(N_a)$ converge to the desired 
frequencies $P_i$, see Eq.~(\ref{C_A}). Substantial 
fluctuations may however occur, for example when 
the agent is matched repeatedly to opponents with 
low levels of expertise, which may lead to an 
extended streak of boring games.
The resulting temporary discrepancy,
\begin{equation}
M_i = P_i-p_i(N_a)\,,
\label{M_i}
\end{equation}
between desired and trailing emotion probabilities 
can then be regarded as an emotional drive. 
Stochastically, $M_i$ averages out for appropriate 
probabilities $q_\alpha$ to select an activity 
$\alpha$. On a shorter time scales one may endorse 
the agent with the option to reduce excessive
values of $M_k$ by direct action, viz by selecting
an activity $\beta =\mbox{Go},\, \mbox{Chess},\, ...$ 
characterized by large/small $p_k^\beta$
when $M_k$ is strongly positive/negative. This is 
meaningful in particular if the distribution $\{p_i^\beta\}$ 
is peaked and not flat. Emotional drives correspond 
to an additional route for reaching the overall 
goal, the alignment of experience with character.

\subsection{Including utility maximization}

In addition to having emotional motivations, 
agents may want to maximize one or more classical 
reward functions, like gaining credits for wining 
games, or answering a substantial number of questions 
in a chat room. Without emotional constraints, the 
program would just select the most advantageous 
option, once the available options have been 
explored in sufficient depth for their properties, 
in analogy to the multi-armed bandit problem 
\cite{vermorel2005multi}. 
An interesting constellation arises when rewards are 
weighted emotionally, e.g.\ with the help of the 
Kullback-Leibler divergence $D_\alpha$ between the 
character and the emotional experience 
of a given behavioral option \cite{gros2015complex},
\begin{equation}
D_\alpha = \sum_i P_i\log\left(\frac{P_i}{p_i^\alpha}\right)\,.
\label{D_alpha}
\end{equation}
Credits received from a behavioral option $\alpha$ 
that conforms with the character of the agent, 
having a small $D_\alpha$, would be given a higher 
weight than credits gained when engaging in 
activities characterized by a large $D_\alpha$.
There are then two conflicting goals, to maximize the
weighted utility and to align experience with character,
for which a suitable prioritization or Pareto optimality
may be established \cite{sener2018multi}.

Instead of treating utility as a separate feature,
one may introduce a new emotional trait, the desire 
to receive rewards, viz to make money, and subsume 
utility under emotional optimization on an equal
footing. Depending on the target frequency $P_U$ 
to generate utility, the agent will select its 
actions such that the full emotional spectrum
is taken into account. A separate weighting of
utility gains, as expressed by (\ref{D_alpha}), is
then not necessary.

\section{Discussion}

Computational models of emotions have focused
traditionally on the interconnection between emotional 
stimuli, synthetic emotions and emotional responses 
\cite{rodriguez2015computational}. A typical 
goal is to generate believable behavior of 
autonomous social agents \cite{scherer2009emotions},
in particular in connection with psychological
theories of emotions, involving f.i.\ appraisal, 
dimensional aspects or hierarchical structures
\cite{rodriguez2015computational}. Closer to the
scope of the present investigation are proposals
relating emotions to learning and with this to behavioral
choices \cite{gadanho2003learning}. One possibility
is to use homeostatic state variables, encoding f.i.\
`well-being', for the regulation of reinforcement 
learning \cite{moerland2018emotion}. Other state 
variables could be derived from utility optimization, 
like water and energy uptake, or appraisal concepts 
\cite{moerland2018emotion}, with the latter being
examples for the abstract evaluation criteria used 
in the TAES framework. One route to measure well-being 
consist in grounding it on the relation between 
short- and long-term trailing reward rates 
\cite{broekens2007affect}. Well-being can then be used 
to modulate dynamically the balance between exploitation 
(when doing well) and exploration (when things are not 
as they used to be). Alternatively, emotional states 
may impact the policy \cite{kuremoto2013improved}.

Going beyond the main trust of research in synthetic
emotions, to facilitate human-computer interaction and
and to use emotions to improve the performance of 
machine learning algorithms that are applied to dynamic 
landscapes, the question that has been asked here 
regards how an ever ongoing sequence of distinct 
tasks can be generated by optimizing emotional 
experience, in addition to reward. Formulated as a 
time allocation problem, the rational of our
approach is drawn mainly from affective neuroscience 
\cite{gros2009emotions}, and only to a lesser extent 
from psychological conceptualizations of 
human emotional responses. Within this setting, 
the TAES framework captures the notion that a 
central role of emotions is to serve as abstract 
evaluation tools that are to be optimized as a set, 
and not individually. This premise does not 
rule out alternative emotional functionalities.

Emotions are considered to be grounded in `affect',
viz in the brain states mediating pleasant and
unpleasant arousal \cite{wilson2013neural}. This
seems at first a contradiction to the notion that 
emotions correspond to `abstract' evaluation 
criteria, as advocated here. It is worthwhile 
to point out in this context that emotions are 
intrinsically related to `domain-general' neural 
processes \cite{barrett2009future,chen2018domain},
and that moral judgments seem to recruit, on
a related note, domain-general 
valuation mechanisms on the basis of probabilistic 
representations \cite{shenhav2010moral}. One
can be frustrated when failing to perform while
playing violin, to illustrate this point,
or when getting a ticket for driving too fast.
Frustration may arise, like any other emotional 
state, in highly diverse domains. In this sense,
domain-general processes and valuation mechanisms
can be termed to be abstract.

\subsection{Testing of functional emotional frameworks}

For living beings, capabilities are
selected ultimately when they contribute
to evolutionary success. This holds in 
particular also for emotional regulation.
A closely related area is the formation 
of moral preferences, an issue that
is examined by a rapidly growing body
of game-theoretical approaches
\cite{capraro2021mathematical}. Engaging
in seemingly unselfish behavior comes 
in this view with personal benefits.
In this context, rational choice theory 
presumes that agents act rationally,
given their personal resource limitations 
and preferences \cite{dietrich2013preferences}.

Classical game-theoretical concepts can be 
tested using suitable laboratory protocols
\cite{camerer2015behavioral}. Evidence
becomes somewhat more indirect when the 
direct maximization of monetary utility
is complemented by personal preferences that
are hypothesized to include moral components, like
fairness and retaliation \cite{fehr2000fairness}.
Testing conceptual frameworks for game-theoretical
settings in which moral preferences are allowed
to evolve is even less straightforward 
\cite{chandan2016rational}. This observation
holds also for the here proposed framework,
TAES, in which preferences, f.i.\ to engage
in challenging tasks, may fluctuate strongly,
being defined only by their long-term average.
Any protocol for testing emotional frameworks
will be bounded by this caveat. 

Detailed testing protocols for TAES are yet
to be developed. They would be based in any
case on a setting, in which participants
are given not one, but several different 
tasks to perform. It would be up to the 
participants to select the relative 
frequency, viz the number times they engage
in any one of the possible tasks. The 
individual tasks would be conceptually
similar, differing however quantitatively
along several feature dimensions. E.g.,
one task could be complex, but 
mildly challenging, another seemingly
simple, but somewhat difficult. In order
to include variability, one could 
include a simple but strongly
varying type of task. The timeline of
task selection would then be compared
with a previously taken character evaluation 
of the participant. The outcome of the 
experiment would be in agreement with TAES 
if character and the statistics of the 
timeline of actions would align.

\section{Conclusions}

The here developed concept, time allocation
via emotional stationarity (TAES), can
be seen from two viewpoints. On one side as
a guiding hypothesis for studies of the brain.
TAES serves in this context as an example 
for the working of emotions in terms of abstract 
evaluation criteria. On the other side, TAES
can be seen as a first step towards the 
implementation of truely synthetic emotions,
viz emotions that mirror human emotion not 
only on a phenomenological, but on a functional level.

Frameworks for synthetic emotions are especially 
powerful and functionally close to human emotions 
if they can be extended with ease along two 
directions. Firstly, as argued in this study, when 
the protocol for the inclusion of new behavioral 
options is applicable to a wide range of activity 
classes.  This is the case when emotions do not 
correspond to specific features, but to domain-general 
evaluation criteria. Essentially any type of activity 
can then be evaluated, f.i., as being boring, 
challenging, risky, demanding, easy, and so on. 
It is also desirable that the framework allows for 
the straightforward inclusion of new traits of 
emotions, such as longing for monetary rewards.

Frameworks for the understanding of the emotional
system should be able to explain that humans 
dispose of characteristic personalities
\cite{deyoung2009personality,mcnaughton2018some}.
For theories of emotions this implies that there 
should exist a restricted set of parameters controlling 
the balancing of emotional states in terms of a 
preferred distribution, the functional equivalent of 
character. As realized by the TAES framework, the 
overarching objective is consequently to adjust the 
relative frequencies to engage in a specific task, 
such that the statistics of the experienced emotional 
states aligns with the character.

Human life is characterized by behavioral options, 
such as to study, to visit a friend, or to take a swim 
in a pool, that have strongly varying properties and
multi-variate reward dimensions. It is hence questionable
whether utility optimization in terms of a univariate
money-like credit, e.g.\ as for the multi-armed bandit
problem, would suffice for an understanding of human
motivational drives. A resolution of this conundrum
is the concept of emotions as domain-general evaluation 
criteria. In this perspective, life-long success depends 
not only on the algorithmic capability to handle specific 
tasks, but also on the alignment of experiences and
personality.

\section*{Acknowledgment}

--none--

\section*{Data availability}

There is no associated data.


\end{document}